\documentclass[12pt]{elsarticle}
\usepackage{refmerge}
\usepackage{epsfig}
\begin{document}
\date{\today}

\title{\bf{ \boldmath
STUDY OF THE PROCESS $e^+e^-\to K_S^0 K_S^0 \pi^+\pi^-$
IN THE C.M. ENERGY RANGE 1.6--2.0 GEV WITH THE CMD-3 DETECTOR
}}

\author[adr1,adr2]{R.R.~Akhmetshin}
\author[adr1,adr2]{A.N.~Amirkhanov}
\author[adr1,adr2]{A.V.~Anisenkov}
\author[adr1,adr2]{V.M.~Aulchenko}
\author[adr1]{V.Sh.~Banzarov}
\author[adr1]{N.S.~Bashtovoy}
\author[adr1,adr2]{D.E.~Berkaev}
\author[adr1,adr2]{A.E.~Bondar}
\author[adr1]{A.V.~Bragin}
\author[adr1,adr2,adr5]{S.I.~Eidelman}
\author[adr1,adr2]{D.A.~Epifanov}
\author[adr1,adr2,adr3]{L.B.~Epshteyn}
\author[adr1,adr2]{A.L.~Erofeev}
\author[adr1,adr2]{G.V.~Fedotovich}
\author[adr1,adr2]{S.E.~Gayazov}
\author[adr6]{F.J. Grancagnolo}
\author[adr1,adr2]{A.A.~Grebenuk}
\author[adr1,adr2]{S.S.~Gribanov}
\author[adr1,adr2,adr3]{D.N.~Grigoriev}
\author[adr1,adr2]{F.V.~Ignatov}
\author[adr1,adr2]{V.L.~Ivanov}
\author[adr1]{S.V.~Karpov}
\author[adr1]{A.S.~Kasaev}
\author[adr1,adr2]{V.F.~Kazanin}
\author[adr1,adr2]{I.A.~Koop}
\author[adr1,adr2]{A.A.~Korobov}
\author[adr1,adr3]{A.N.~Kozyrev}
\author[adr1,adr2]{E.A.~Kozyrev}
\author[adr1,adr2]{P.P.~Krokovny}
\author[adr1,adr2]{A.E.~Kuzmenko}
\author[adr1,adr2]{A.S.~Kuzmin}
\author[adr1,adr2]{I.B.~Logashenko}
\author[adr1,adr2]{P.A.~Lukin}
\author[adr1]{A.P.~Lysenko}
\author[adr1,adr2]{K.Yu.~Mikhailov}
\author[adr1]{V.S.~Okhapkin}
\author[adr1,adr2]{E.A.~Perevedentsev}
\author[adr1]{Yu.N.~Pestov}
\author[adr1,adr2]{A.S.~Popov}
\author[adr1,adr2]{G.P.~Razuvaev}
\author[adr1]{A.A.~Ruban}
\author[adr1]{N.M.~Ryskulov}
\author[adr1,adr2]{A.E.~Ryzhenenkov}
\author[adr1,adr2]{A.V.~Semenov}
\author[adr1]{Yu.M.~Shatunov}
\author[adr1,adr2]{V.E.~Shebalin}
\author[adr1,adr2]{D.N.~Shemyakin}
\author[adr1,adr2]{B.A.~Shwartz}
\author[adr1,adr2]{D.B.~Shwartz}
\author[adr1,adr4]{A.L.~Sibidanov}
\author[adr1,adr2]{E.P.~Solodov\fnref{tnot}}
\author[adr1]{M.V.~Timoshenko}
\author[adr1]{V.M.~Titov}
\author[adr1,adr2]{A.A.~Talyshev}
\author[adr1]{A.I.~Vorobiov}
\author[adr1]{I.M.~Zemlyansky}
\author[adr1,adr2]{Yu.V.~Yudin}

\address[adr1]{Budker Institute of Nuclear Physics, SB RAS, 
Novosibirsk, 630090, Russia}
\address[adr2]{Novosibirsk State University, Novosibirsk, 630090, Russia}
\address[adr3]{Novosibirsk State Technical University, 
Novosibirsk, 630092, Russia}
\address[adr4]{University of Victoria, Victoria, BC, Canada V8W 3P6}
\address[adr5]{Lebedev Physical Institute RAS, Moscow, 119333, Russia}
\address[adr6]{Instituto Nazionale di Fisica Nucleare, Sezione di Lecce, Lecce, Italy}
\fntext[tnot]{Corresponding author: solodov@inp.nsk.su}


%
\vspace{0.7cm}
\begin{abstract}
\hspace*{\parindent}
The cross section of the process $e^+e^- \to K_S^0 K_S^0\pi^+\pi^-$ has been 
measured using a data sample of 56.7 pb$^{-1}$ collected 
with the CMD-3 detector at the VEPP-2000  $e^+e^-$ collider.  596$\pm$27 
and 210$\pm$18 signal events have been selected with six and five detected 
tracks, respectively,  in the center-of-mass energy range 
1.6--2.0 GeV. The total systematic uncertainty of the cross section is
about 10\%. The study of the production dynamics confirms
the dominance of the  $K^*(892)^+ K^*(892)^-$  intermediate state. 
\end{abstract}

\maketitle
\baselineskip=17pt
\section{ \boldmath Introduction}
\hspace*{\parindent}
$e^+e^-$ annihilation into hadrons below 2 GeV is rich for various
multiparticle final states. Their detailed studies are important for
development of phenomenological models describing strong interactions
at low energies. 
One of the final states, $ K_S^0 K_S^0\pi^+\pi^-$, 
has been studied before by the BaBar collaboration~\cite{isrksks2pi},
based on the Initial-State Radiation (ISR) method. 
Their analysis showed that below the center-of-mass energy ($E_{\rm c.m.}$) 
of 2 GeV  the process is dominated by the  $K^*(892)^+ K^*(892)^-$  
intermediate state with a small contribution of the $K_S^0 K_S^0\rho(770)$ 
reaction.
As a part of the total hadronic cross section, the cross section of  
$e^+e^- \to K_S^0 K_S^0\pi^+\pi^-$ is interesting for   
the calculations of the hadronic vacuum polarization and, as a consequence,
for the hadronic contribution to the muon anomalous magnetic 
moment~\cite{g21,g22,g23}. Until recently, of various possible charge
combinations of the $K \bar K\pi\pi$ final state only two were 
measured ($K^+K^-\pi^+\pi^-$ and $K^+K^-\pi^0\pi^0$). Contributions from 
other $K \bar K\pi\pi$ final states ($K^{\pm}K^0_S\pi^\mp\pi^0$,
$K^0_SK^0_S\pi^0\pi^0$ etc.) were taken into account using isospin relations
that resulted in large uncertainties. The measurements of other exclusive
reactions, see~\cite{TheBABAR:2017aph} and references therein, helped   
decreasing such uncertainties and changed the contribution of such final states
to the muon anomalous magnetic moment from $3.31\pm0.58$ to $2.41\pm0.11$ in units
of $10^{-10}$ for the energy range below 2 GeV.
The difference is rather large and the detailed study of the production 
dynamics can further improve the accuracy of these calculations and 
understanding of the energy dependence of the cross section.

In this paper we report the analysis of the data sample of 
56.7 pb$^{-1}$ collected at the CMD-3 detector
in the 1.6--2.0 GeV $E_{\rm c.m.}$ range. 
These data were collected during four energy scans, with a 5--10 MeV  
c.m. energy step each, performed  at  the VEPP-2000 $e^+e^-$ 
collider~\cite{vepp1,vepp2,vepp3,vepp4}
in the 2011, 2012 and 2017  experimental runs. 
In 2017 (about half of integrated luminosity) the beam energy was monitored by 
the back-scattering laser-light system~\cite{laser1,laser2}, providing
an absolute beam-energy monitoring with better than 0.1 MeV uncertainty 
at every 10-20 minutes of data taking. In earlier runs the beam energy 
was determined using measurements of charged track momenta in the 
detector magnetic field with an about 1 MeV uncertainty.
Since the cross section of the process is small, we combine our scanned 
points into eight energy intervals as shown in Table~\ref{xs_all}.

The general-purpose detector CMD-3 has been described in 
detail elsewhere~\cite{sndcmd3}. Its tracking system consists of a 
cylindrical drift chamber (DC)~\cite{dc} and double-layer multiwire 
proportional 
Z-chamber, both also used for a charged track trigger, and both inside 
a thin (0.2~X$_0$) superconducting solenoid with a field of 1.3~T.
The tracking system provides the 98-99\% tracking efficiency in about 
70\% of the solid angle.
The liquid xenon (LXe) barrel calorimeter with a 5.4~X$_0$ thickness has
fine electrode structure, providing 1--2 mm spatial resolution for photons 
independently of their energy~\cite{lxe}, and is located   
in the cryostat vacuum volume outside the solenoid.     
The barrel CsI crystal calorimeter with a thickness 
of 8.1~X$_0$ surrounds the LXe calorimeter, while the endcap BGO 
calorimeter with a thickness of 13.4~X$_0$ is placed inside the 
solenoid~\cite{cal}.
Altogether, the calorimeters cover 0.9 of the solid angle and
amplitude signals provide information for the neutral trigger.
Charged trigger requires presence of only one charged track in DC,
therefore it has practicaly 100\% efficiency for our studied process 
with five  or six detected tracks. A relatively large fraction of
these events has sufficient energy deposition in the calorimeter for 
the independent neutral trigger: these events are used to control 
the charged trigger efficiency. The luminosity is measured using the 
Bhabha scattering events at large angles with about 1\% systematic 
uncertainty~\cite{lum}. 

To understand the detector response to processes under study and to
obtain the detection efficiency, we have developed Monte Carlo (MC) 
simulation of our detector based on the GEANT4~\cite{geant4} package, 
in which all simulated events pass the whole reconstruction and selection 
procedure. The MC simulation uses primary generators with the matrix elements 
for the  $K_S^0 K_S^0\pi^+\pi^-$ final state with the $K^*(892)^+ K^*(892)^-$, 
$K_1(1400)K_S^0\to K^*(892)^{\pm}\pi^{\mp} K_S^0$, 
 and $K_1(1270)K_S^0\to  K_S^0\rho(770) K_S^0$ intermediate states. The 
primary generator with the  $K_S^0 K_S^0\pi^+\pi^-$ in the phase-space model 
(PS) has been also developed.
The primary generator includes radiation of photons by an initial 
electron or positron, calculated according to Ref.~\cite{kur_fad}. 
\section{Selection of $e^+e^-\to K_S^0 K_S^0\pi^+\pi^-$ events}
\label{select}
\hspace*{\parindent}

The analysis procedure is similar to our study of the production of
six charged pions described in Ref.~\cite{cmd6pi}. 
Candidate events
are required to have five or six charged-particle tracks, each having: 
\begin{itemize}
\item{
more than five hits in the DC;
}
\item{
a transverse momentum larger than 40 MeV/c;
}
\item{
a minimum distance from a track to the beam axis in the
transverse  plane of less than 6 cm, that allows reconstruction of a decay 
point of $K_S^0$ at large distances;
}
\item{
a minimum distance from a track to the center of the interaction region along
the beam axis Z of less than 15 cm.
}
\end{itemize} 

Reconstructed momenta and angles of the tracks for the five- and six-track 
events are used for further selection. 
  
In our reconstruction procedure we create the list of the $K_S^0\to\pi^+\pi^-$ 
candidates which includes every pair of oppositely charged tracks, assuming 
them to be pions, with the invariant mass within $\pm 80$ Mev/c$^2$ from 
the $K_S^0$ mass~\cite{pdg}
and a common vertex point within a spacial uncertainty of the DC.  
We calculate momentum and energy for each $K_S^0$ candidate taking 
the value of the $K_S^0$ mass from Ref.~\cite{pdg}. 

At the first stage of signal event selection we require at least two $K_S^0$ 
candidates with four independent tracks plus one  or two additional charged 
tracks originating from the collision point. 
If there are still more than two $K_S^0$ candidates, two candidate pairs with 
minimal deviations from the $K_S^0$ mass are retained.  Additionally, we 
require the distance from the 
beam axis for the tracks not from $K_S^0$ to be less than 0.35 cm.

\begin{figure}[tbh]
\begin{center}
\vspace{-0.cm}
\includegraphics[width=1.0\textwidth]{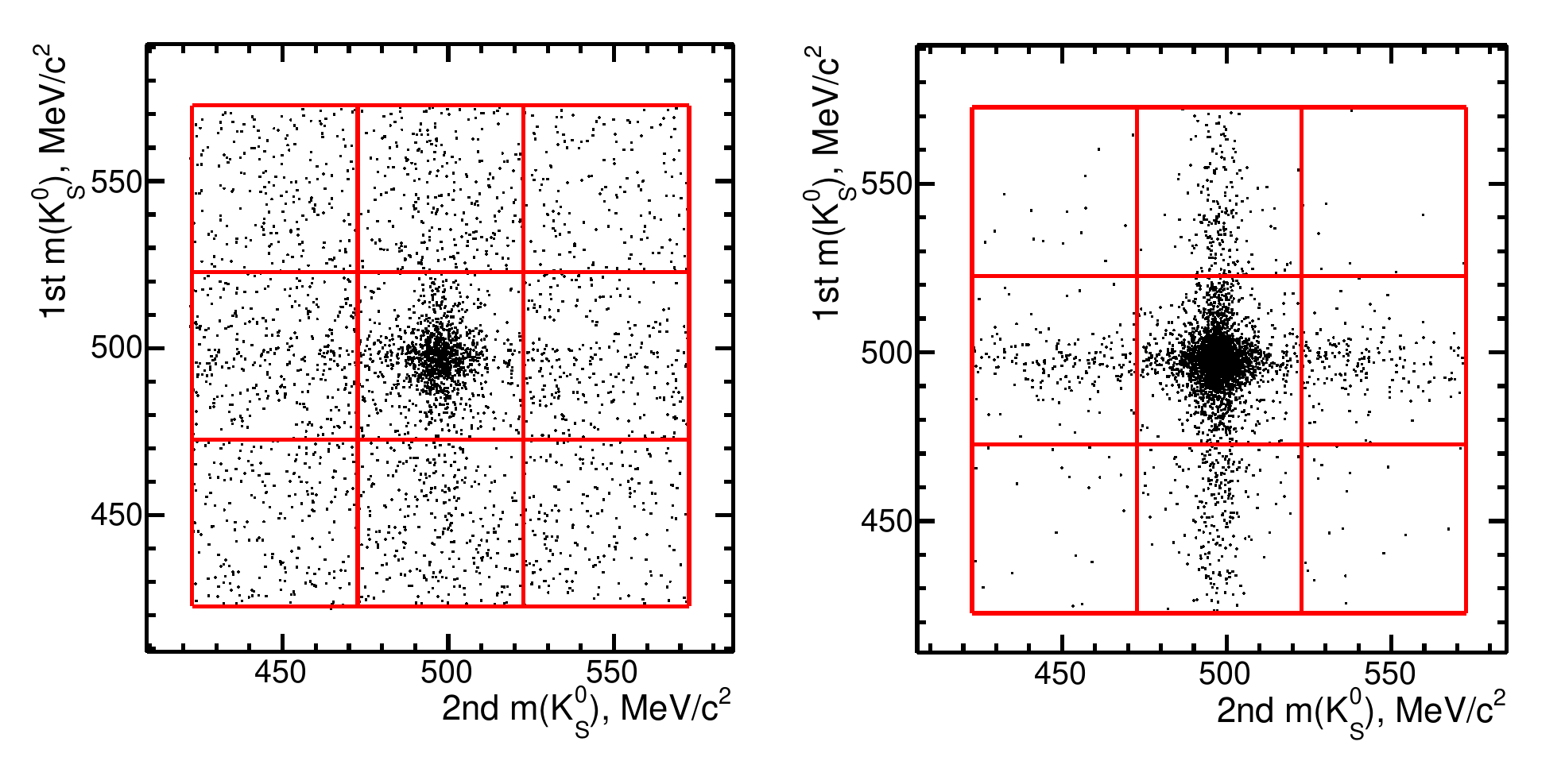}
\put(-245,170){\makebox(0,0)[lb]{\bf(a)}}
\put(-45,170){\makebox(0,0)[lb]{\bf(b)}}
\vspace{-0.8cm}
\caption
{
Scatter plot of the invariant mass of one
 $K_S^0 $ candidate vs  invariant mass for the second $K_S^0 $ candidate 
for data (a) and MC simulation (b).
The lines show selections for the signal events and for the background level 
estimate.
}
\label{m1vsm2}
\end{center}
\end{figure}

Figure~\ref{m1vsm2} shows the scatter plot for the invariant mass of the first 
$K_S^0\to\pi^+\pi^-$ candidate vs the second one for data (a) and MC 
simulation (b). All energy intervals are combined for the presented histograms 
in data. The lines show selections for the signal events in the central square 
and for the background level estimate from the events in other eight squares. 

\begin{figure}[tbh]
\begin{center}
\vspace{-0.cm}
\includegraphics[width=1.0\textwidth]{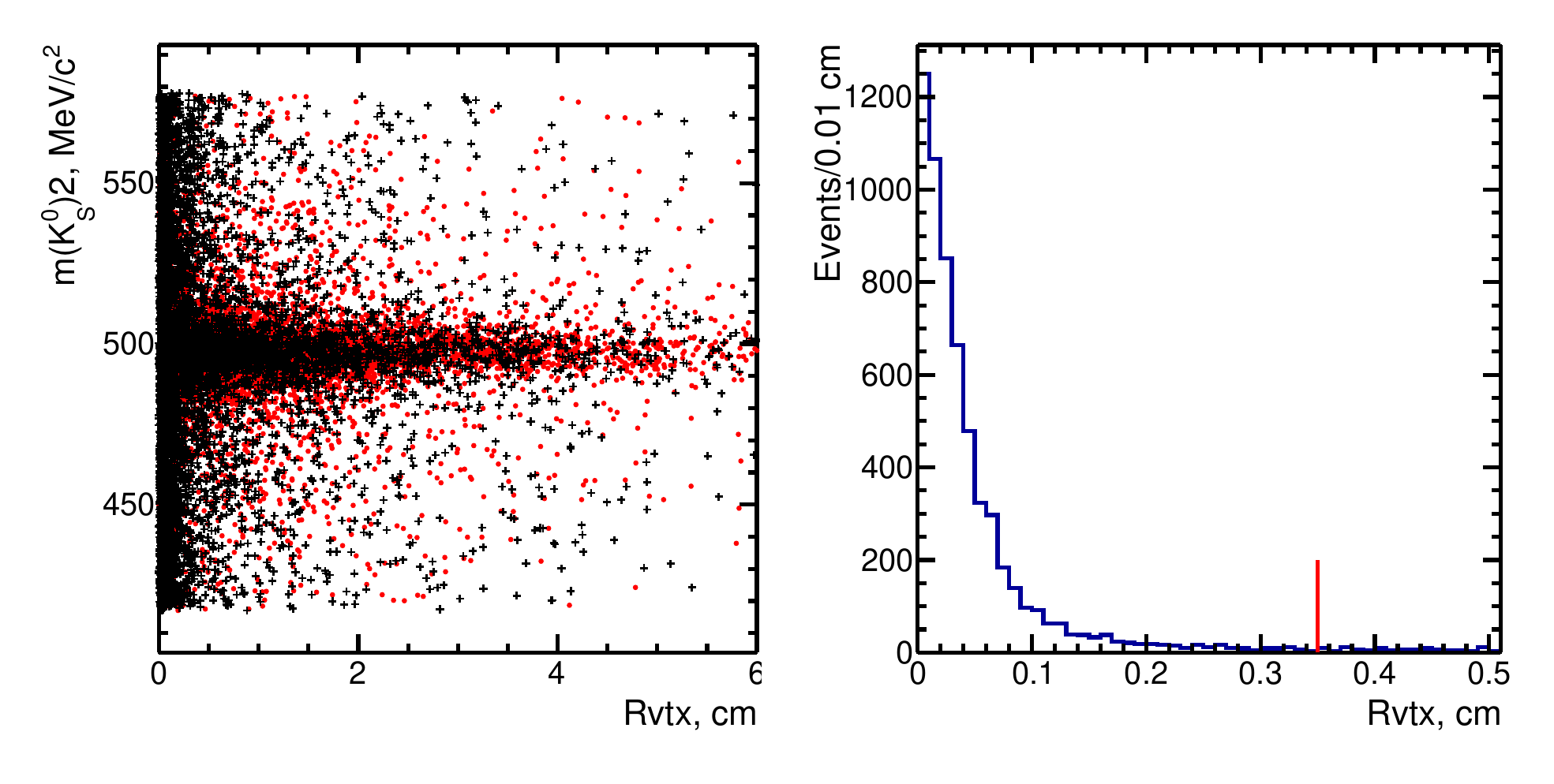}
\put(-245,170){\makebox(0,0)[lb]{\bf(a)}}
\put(-45,170){\makebox(0,0)[lb]{\bf(b)}}
\vspace{-0.8cm}
\caption
{
(a) Scatter plot of the invariant mass of the
 $K_S^0 $ candidate vs the radial distance of the decay vertex (crosses): 
MC simulation is shown by (red in color version) dots. 
(b) Radial distance of the not-from-kaon pions from the beam axis.
The line shows applied selection.
}
\label{mksvsrad}
\end{center}
\end{figure}

Figure~\ref{mksvsrad}(a) shows the scatter plot for the invariant mass 
of the $K_S^0\to\pi^+\pi^-$ candidates vs the radial distance of the 
reconstructed vertex from the beam axis. Events associated with $K_S^0$ are 
clearly seen as well as the background events. Red dots (in the color version) 
are for simulation. Figure~\ref{mksvsrad}(b) shows a radial distance from the 
beam axis for the tracks not associated with the $K_S^0$ decay. The 
additional requirement for this distance to be less than 0.35 cm is shown 
by the line.

\begin{figure}[tbh]
\begin{center}
\vspace{-0.cm}
\includegraphics[width=1.0\textwidth]{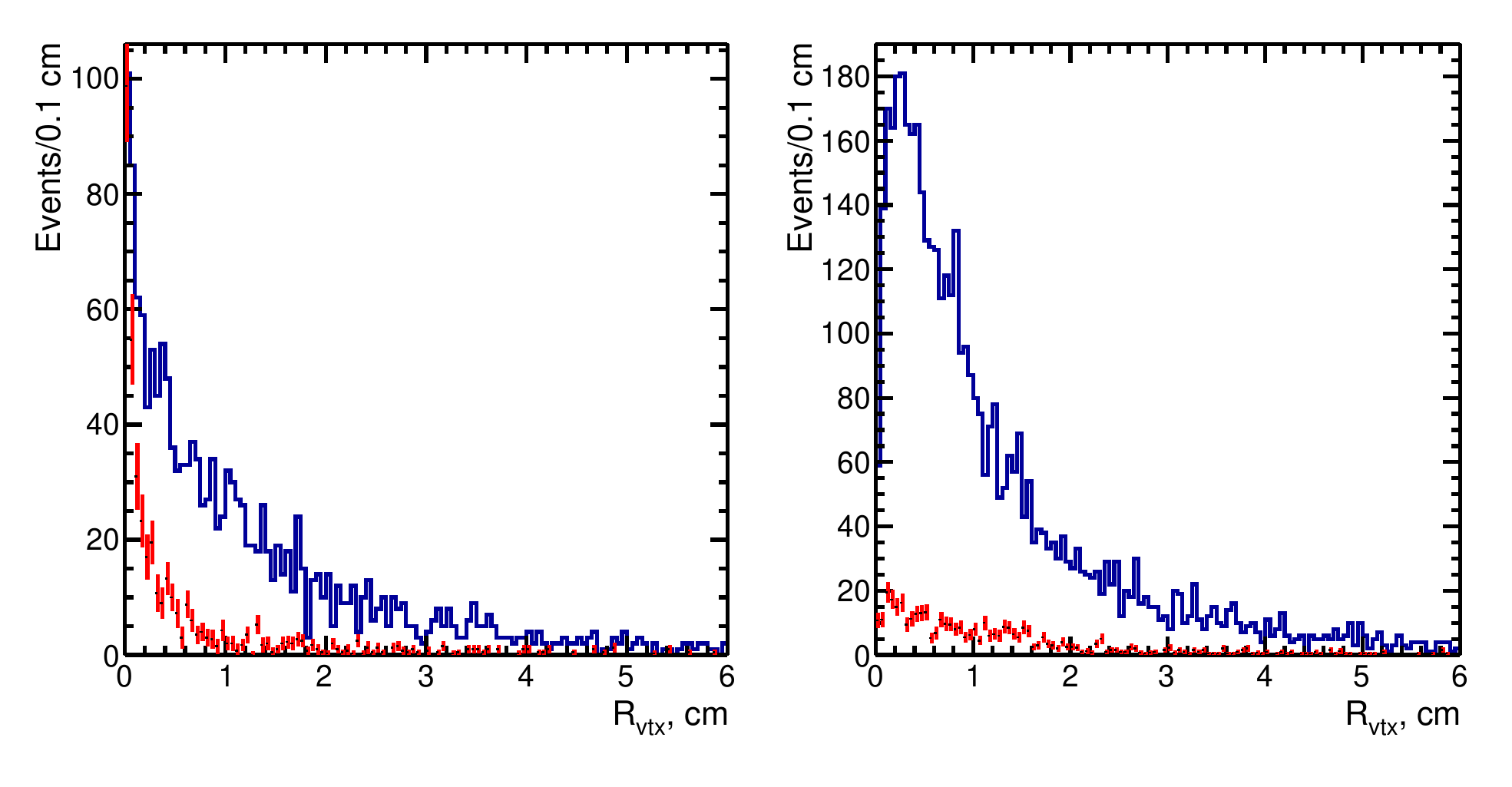}
\put(-245,173){\makebox(0,0)[lb]{\bf(a)}}
\put(-45,173){\makebox(0,0)[lb]{\bf(b)}}
\vspace{-0.8cm}
\caption
{Radial distance of the decay vertex for the $K_S^0\to\pi^+\pi^-$ candidates 
in the signal region of Fig.~\ref{m1vsm2} for data (a) and simulation (b). 
Dots with errors represent the background contribution estimated from 
Eq.~\ref{eqbkg}.
}
\label{ksrad}
\end{center}
\end{figure}
The central vertical and horizontal bands in Fig.~\ref{m1vsm2}
correspond to the 
events with one wrongly reconstructed $K_S^0$, which are seen in 
Fig.~\ref{m1vsm2}(b) also for simulation due to a combinatorial effect, or if 
one of the $K_S^0$ decays to other modes. For data these events can be also 
due to a possible background, like $e^+e^-\to K_S^0K^{\pm}\pi^{\mp}\pi^+\pi^-$ 
with a misidentified or missing charged kaon, when only one $K_S^0$ 
is present in the 
final state: these events contribute to the selected data sample in 
Fig.~\ref{m1vsm2}(a). 

While we do not use special ordering for the calculation of the
first and second $m(K_S^0)$, the background contribution to the
vertical and horizontal bands can be different, and 
we use the ``nine tile'' method to extract the number of signal events and 
estimate the background contribution. In this procedure the two-dimensional 
plot is divided into nine tiles with equal area as shown in Fig.~\ref{m1vsm2} 
by lines. The tiles in Fig.~\ref{m1vsm2} are numbered from left to right from 
top to bottom.  The central (signal) tile contains  $N_5$ events with two 
well-reconstructed $K_S^0$ candidates, while the
vertical and horizontal tiles, connected to the central signal tile,
are used for an estimate of the background contribution to $N_5$ from the 
wrongly reconstructed and single $K_S^0$ events. The four corner tiles are 
used to estimate the random background. The number of  background events
in the central tile, $N_{\rm bkg}$, is thus determined as 
\begin{equation}
N_{\rm bkg} = (N_{2} + N_{4} +N_{6} + N_{8})/2 - (N_{1} + N_{3} +N_{7} + N_{9})/4.
\label{eqbkg}
\end{equation}
Note that the random background for a single tile is taken twice from the 
vertical and horizontal tiles, so the average random background, estimated 
from the corner tiles, is used for compensation. 

At the next stage of event selection, we calculate the expected 
distribution of any kinematic quantity by weighting the contribution of the
eight tiles as in Eq.~\ref{eqbkg}. This is compared to the  
distribution observed in the signal region.

Figure~\ref{ksrad} shows the histograms for the radial distance of the decay 
vertex for the $K_S^0\to\pi^+\pi^-$ candidates in the signal region of 
Fig.~\ref{m1vsm2} for data (a)  and simulation at 1900 MeV (b). Points 
with errors represent the contribution of the background, estimated by  
the ``nine tile'' method of Eq.~\ref{eqbkg}. The background from the 
beam-originating events in data is 
seen in a few first bins and is well estimated by the method: 
it is not dominating, 
so we do not impose any restrictions on this distance. 
The procedure is also applied to the simulation, and Eq.~\ref{eqbkg} gives 
about 5\% of the ``background'' events (Fig.~\ref{ksrad}(b)) due to only one 
correctly reconstructed $K_S^0$ or due to small non-linearity of events in 
the bands, which is assumed to be linear for the method. In our analysis
these events are treated in the same way as for data. The systematic 
uncertainties of the method are discussed below.

\begin{figure}[tbh]
\begin{center}
\vspace{-0.cm}
\includegraphics[width=1.0\textwidth]{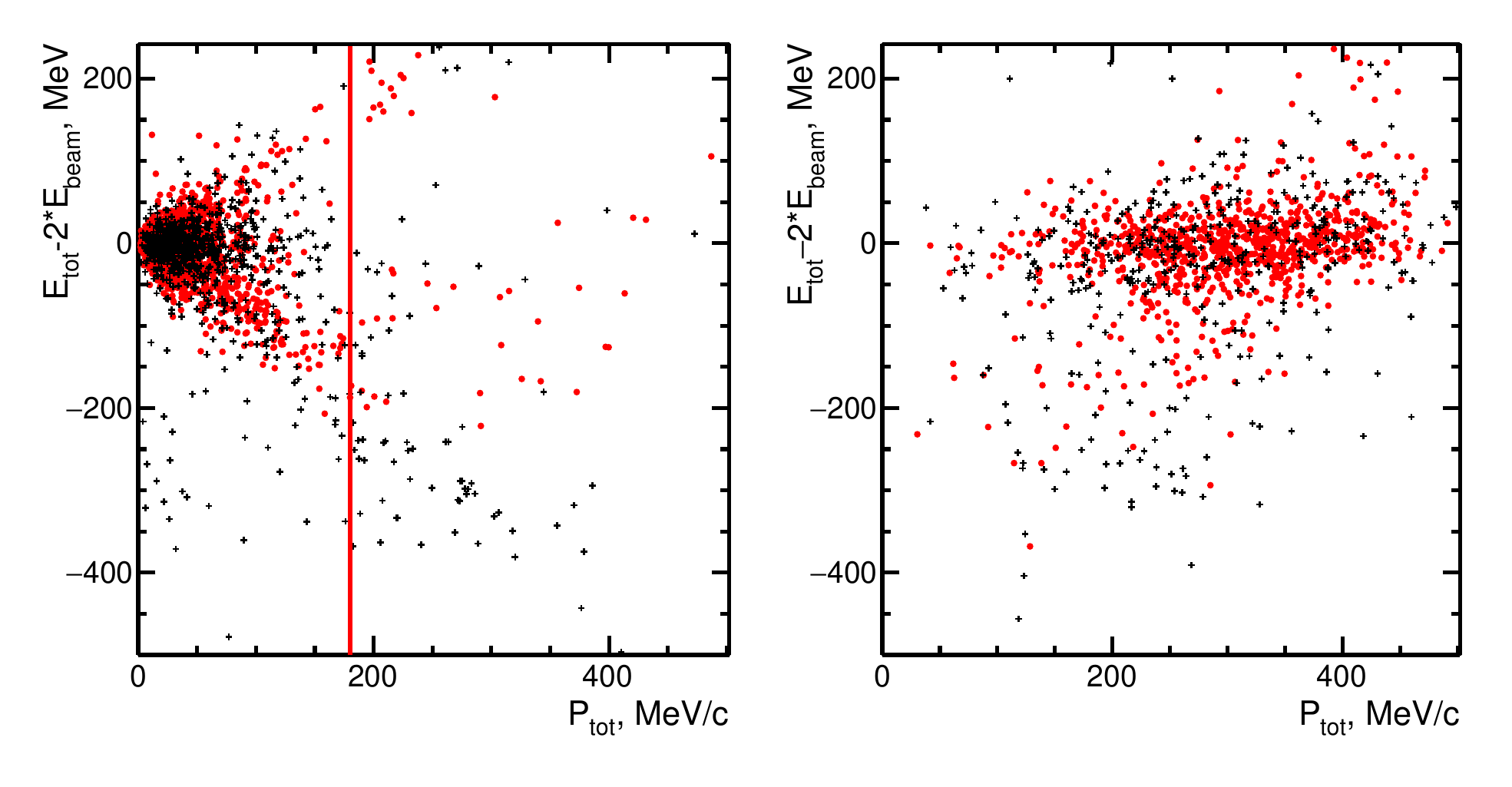}
\put(-245,163){\makebox(0,0)[lb]{\bf(a)}}
\put(-145,163){\makebox(0,0)[lb]{\bf(b)}}
\vspace{-0.8cm}
\caption
{
(a) Scatter plot of the difference  between the energy of 
$K_S^0 K_S^0\pi^+\pi^-$ candidates 
and c.m. energy vs total momentum for events with six tracks. The 
crosses are for data, while the signal simulation is shown by red 
(in color version) dots; the line shows the applied selection.
(b) Scatter plot of the difference  between the energy of 
$K_S^0 K_S^0\pi^+\pi^-$ candidates 
and c.m. energy vs total momentum for events with five tracks. The 
crosses are for data, while the signal simulation is shown by red 
(in color version) dots.
}
\label{energy}
\end{center}
\end{figure}

For the six- or five-track $K_S^0 K_S^0\pi^+\pi^-$ candidates we calculate the
total energy of two $K_S^0$'s and two pions: for the five-track candidates the 
missing  momentum is used to calculate the energy of the lost pion.
Figure~\ref{energy}(a) shows the scatter plot of the difference between 
the total energy and c.m. energy, $\rm E_{tot}-E_{c.m.}$, vs the total
momentum of six- (a) or five-track (b) candidates, $P_{\rm tot}$ for data. 
Events from the central tile  of Fig.~\ref{m1vsm2} are shown.
A clear signal of the $e^+ e^- \to K_S^0 K_S^0\pi^+\pi^-$ reaction is seen in 
Fig.~\ref{energy}(a) as a cluster of crosses near zero, in agreement with  
the expectation from the simulation shown by (red in the color version) dots.
We require $P_{\rm tot}$ to be less than 180 MeV/c$^2$, thus reducing the number of 
events with hard radiative photons.

The expected  signal of five-track candidates has the $\rm E_{tot}-E_{c.m.}$ 
value near zero, and the $\rm P_{tot}$ value is distributed up to 500 MeV/c, 
as shown by the (red) dots from the signal MC simulation in 
Fig.~\ref{energy}(b). The  
(black) crosses show our data: signal events are clearly seen.

\begin{figure}[p]
\begin{center}
\vspace{-0.cm}
\includegraphics[width=1.0\textwidth]{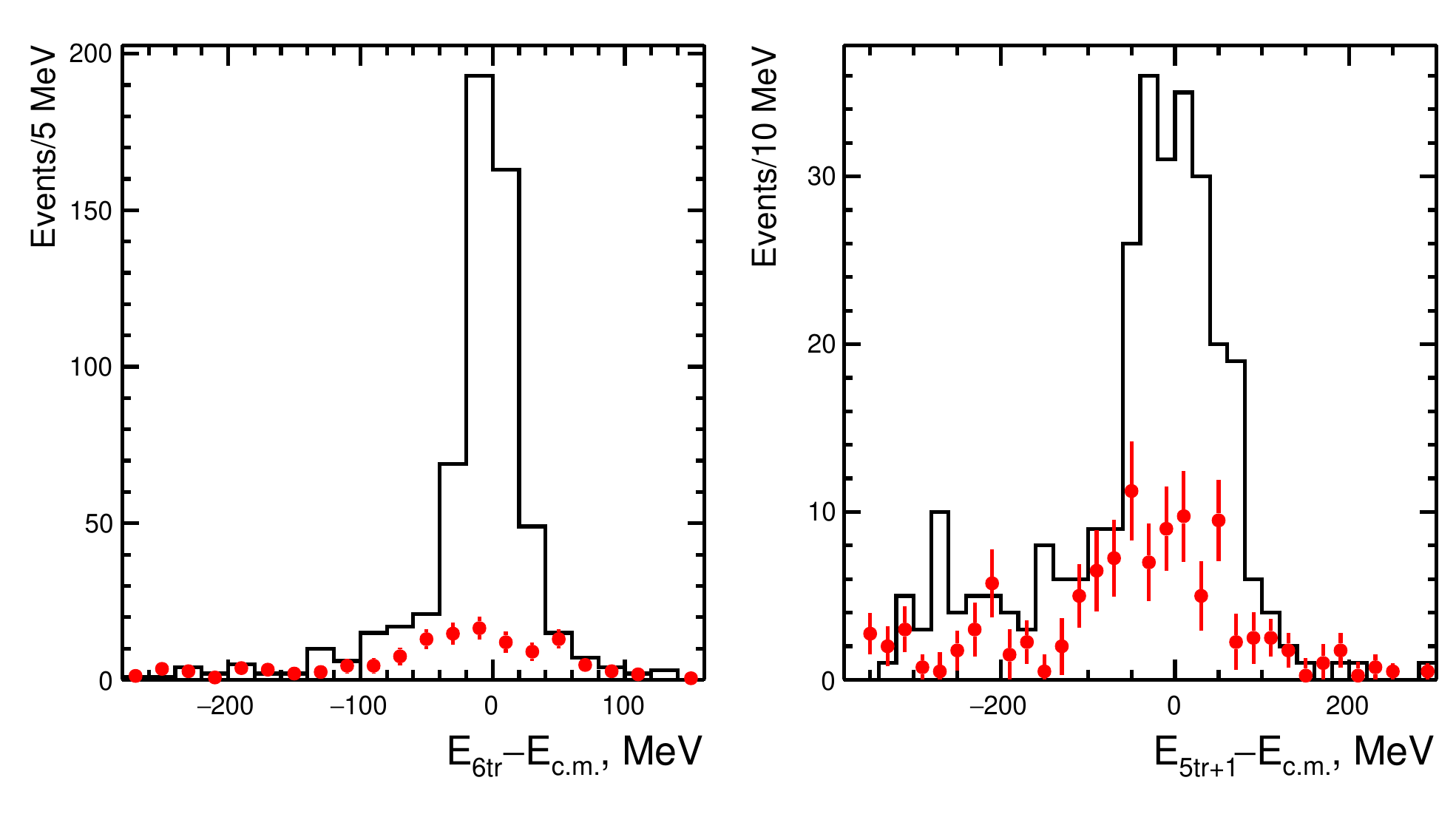}
\put(-245,163){\makebox(0,0)[lb]{\bf(a)}}
\put(-45,163){\makebox(0,0)[lb]{\bf(b)}}
\vspace{-0.8cm}
\caption
{
(a) The difference  between the energy of the $K_S^0 K_S^0\pi^+\pi^-$ 
candidates and c.m. energy after selection by the line in Fig.~\ref{energy} 
for six-track events (a) and five-track events (b). All the energy intervals 
are summed. The dots with errors show the background contribution.
}
\label{energy1D}
\end{center}
%
%
\begin{center}
\vspace{-0.cm}
\includegraphics[width=0.49\textwidth]{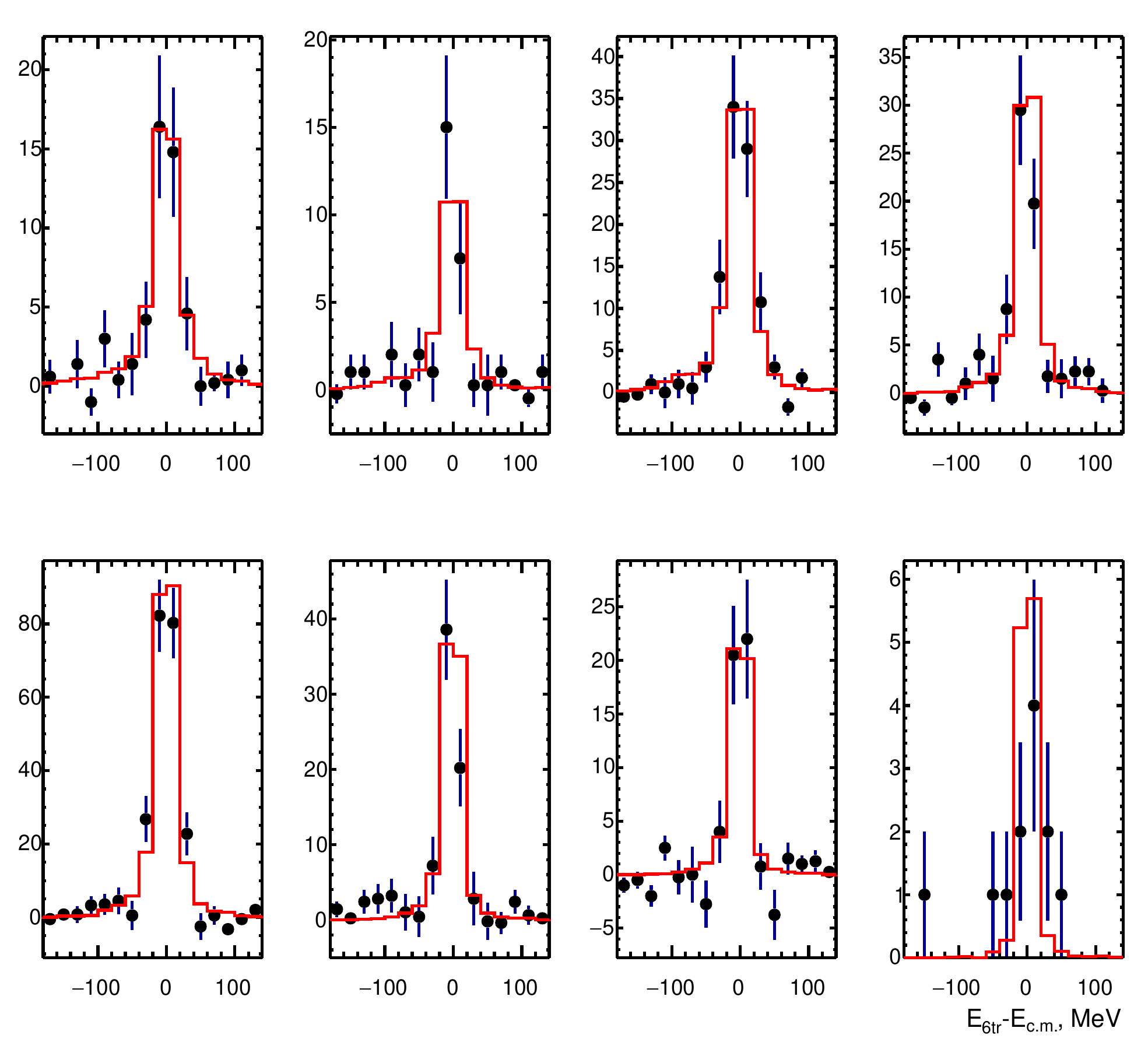}
\hfill
\includegraphics[width=0.49\textwidth]{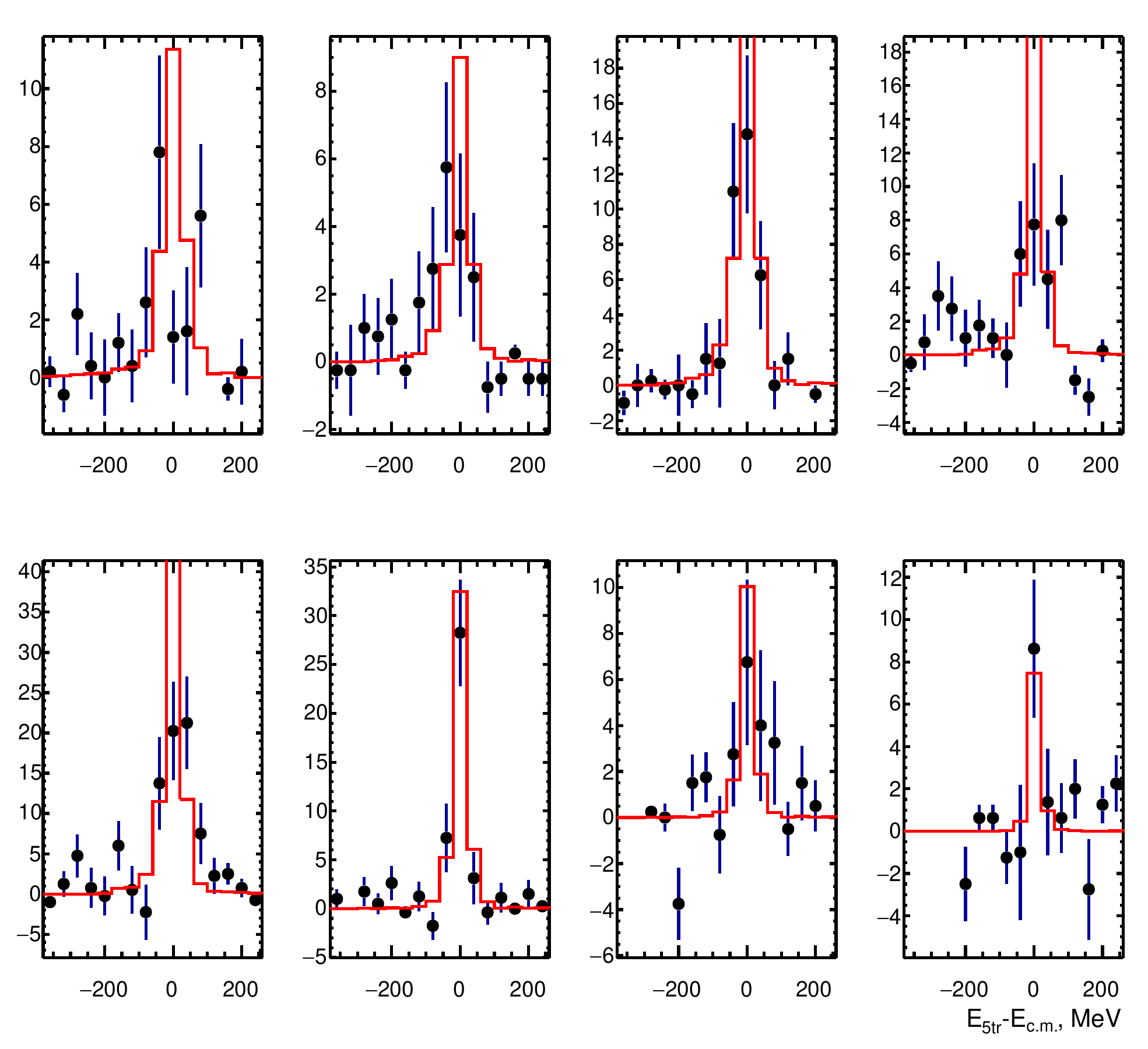}
\put(-245,173){\makebox(0,0)[lb]{\bf(a)}}
\put(-45,173){\makebox(0,0)[lb]{\bf(b)}}
\vspace{-0.2cm}
\caption
{
(a) The difference  between the energy of the $K_S^0 K_S^0\pi^+\pi^-$ 
candidates
and c.m. energy after background subtraction for six-track  (a) and five-track 
events (b) for eight c.m. energy intervals (dots): left to right, 
top to bottom according to Table~\ref{xs_all}. 
Histograms show expected signals from simulation, normalized to the 
total number of events in each plot.
}
\label{6energy}
\end{center}
\end{figure}

Figure~\ref{energy1D} shows the projection plots of Fig.~\ref{energy}, 
$\rm E_{tot}-E_{c.m.}$, for the six-track (a) with applied selection, 
and the five-track (b) events: the histograms present events from 
the signal tile, while dots with errors are our estimate of the background 
contribution using Eq.~\ref{eqbkg}. All energy intervals are summed. 

%

To obtain the number of signal events, we use distributions of 
Fig.~\ref{energy1D} for the $\rm E_{tot}-E_{c.m.}$ difference, i.e. 
$N_{\rm sig} = N_5 - N_{\rm bkg}$.
We subtract the estimated background  
for each energy interval for six- and five-track events, and count remaining 
events in the $\pm$100 MeV region for the six-track events, and in the 
$\pm$200 MeV region for the five-track events. The obtained differences are 
shown in Fig.~\ref{6energy} by dots for six- (a) and five-track (b) events: 
from left to right, from top to bottom according to energy intervals of 
Table~\ref{xs_all}. The histograms show expected signals from the simulation. 
In total, we obtain 596$\pm$27 and 210$\pm$18 for six- and 
five-track signal events, respectively. 
The numbers of selected events determined in each energy interval are 
listed in Table~\ref{xs_all}.
\section{Study of the production dynamics}
\label{dynamics}
\hspace*{\parindent}
First attempts to study the dynamics of the process 
$e^+e^-\to K_S^0 K_S^0\pi^+\pi^-$ were carried out by  
the BaBar Collaboration~\cite{isrksks2pi}. They
reported the observation of the $e^+e^-\to K^*(892)^+K^*(892)^-, ~K^*(892)^{\pm}\pi^{\mp}K_S^0, ~K_S^0 K_S^0\rho(770)$ processes, which contribute to the 
$K_S^0 K_S^0\pi^+\pi^-$ final state, with a dominant production of the 
$K^*(892)^+K^*(892)^-$ intermediate state for the c.m. energies below 2.5 GeV.
In the model of the $K \bar{K}\pi^+\pi^-$ production the  $K_1(1400)K_S^0$ 
intermediate state decays to $K^*(892)^{\pm}\pi^{\mp}K_S^0$, while
$K_1(1270)K_S^0$ can be observed in the $K_S^0 K_S^0\rho(770)$ state.

\begin{center}
\begin{figure}[p]
\vspace{-0.cm}
\includegraphics[width=1.0\textwidth]{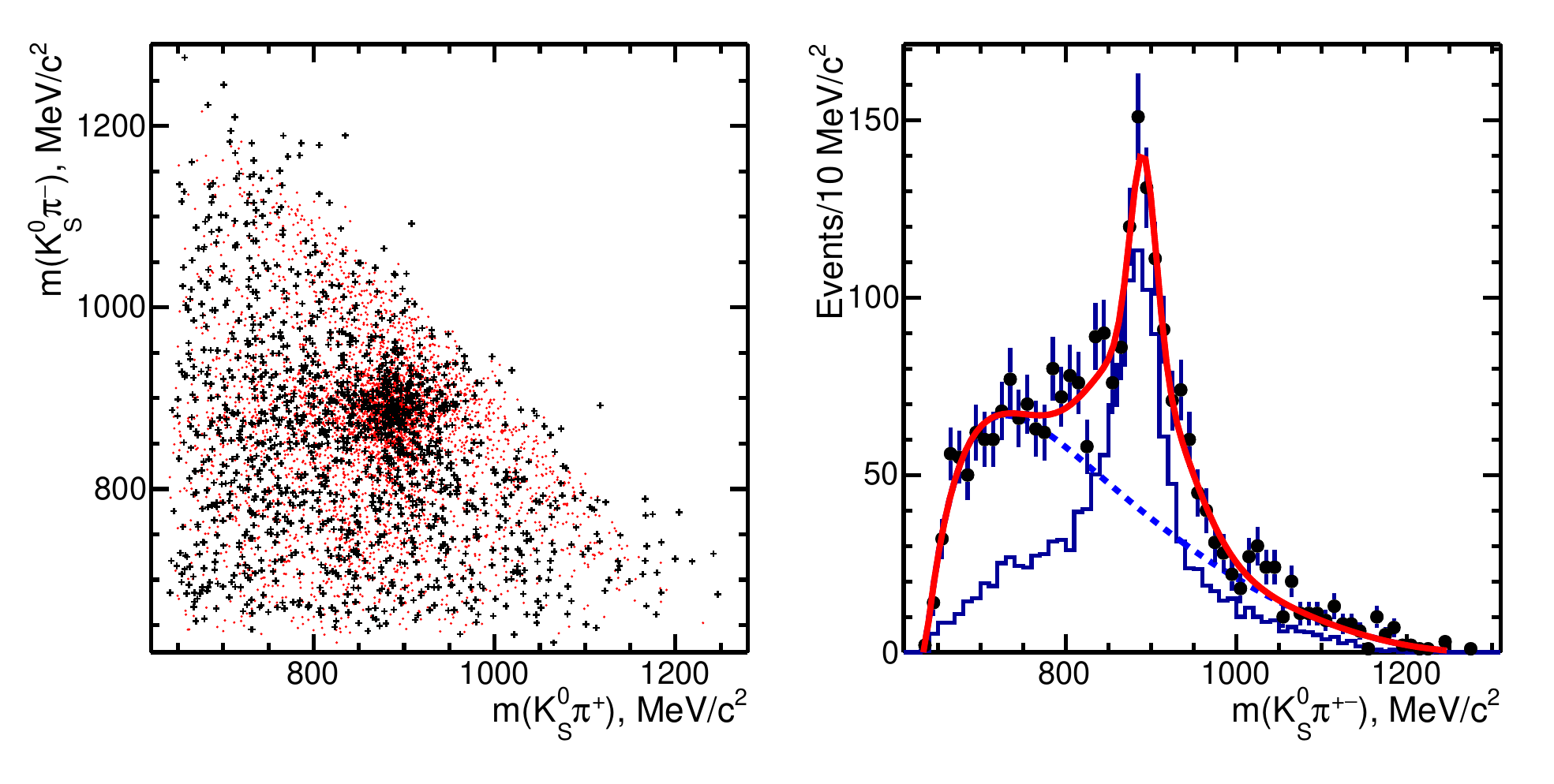}
\put(-245,163){\makebox(0,0)[lb]{\bf(a)}}
\put(-45,163){\makebox(0,0)[lb]{\bf(b)}}
\vspace{-0.5cm}
\caption
{(a) Experimental $K_S^0\pi^-$  vs $K_S^0\pi^+$ invariant mass distribution 
(two entries per event, crosses) for the events from the signal region of 
Fig.~\ref{m1vsm2}. Dots (red in color version) show the simulated 
distribution for the $K^*(892)^+K^*(892)^-$ intermediate state.
(b) Projection plot of (a) (four entries per event) with the fit 
function (solid curve) to determine the number of events with the 
$K^*(892)$ signal over 
the background distribution (dotted curve). The histogram shows the 
corresponding 
number of simulated events for the  $K^*(892)^+K^*(892)^-$ intermediate state.
}
\label{masses1}
%
\includegraphics[width=1.0\textwidth]{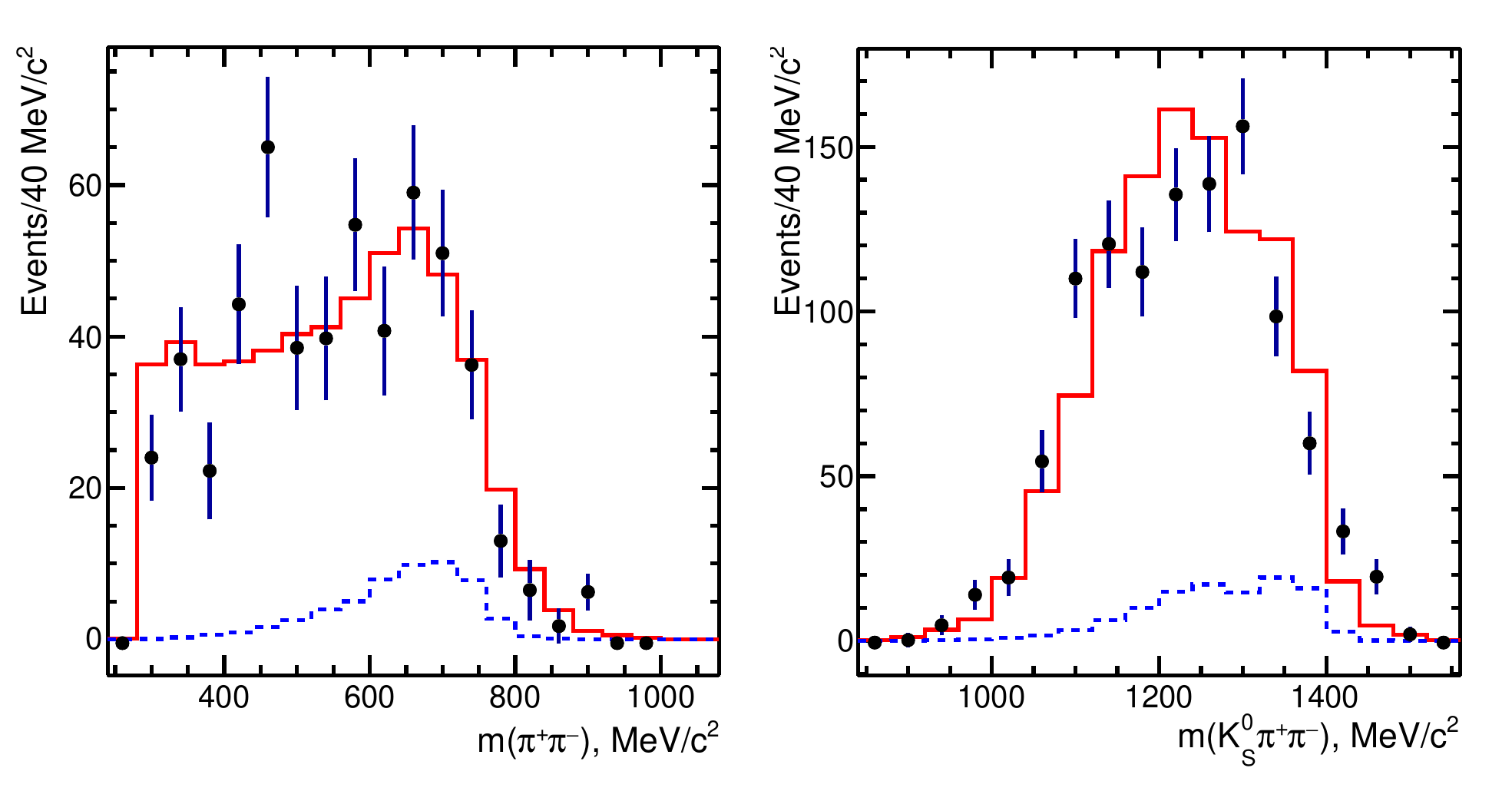}
\put(-245,163){\makebox(0,0)[lb]{\bf(a)}}
\put(-45,163){\makebox(0,0)[lb]{\bf(b)}}
\vspace{-0.5cm}
\caption
{(a) The background-subtracted $\pi^+\pi^-$ invariant mass distribution (dots)
in comparison with the simulated distribution (solid histogram). 
(b) The background-subtracted $K_S^0\pi^+\pi^-$ (two entries per event) 
invariant mass distribution (dots) in comparison with the simulated one 
(a solid histogram).
In both plots the simulated distribution includes the $K^*(892)^+K^*(892)^-$ 
intermediate state plus 30\% of the $K_1(1270)K_S^0$, which contribution is 
shown by a dotted histogram.
}
\label{masses2}
\end{figure}
\end{center}

Figure~\ref{masses1}(a) shows the scatter plot of the $K_S^0\pi^-$ invariant 
mass vs $K_S^0\pi^+$ (two entries/event, all energy intervals are summed) 
for the $K_S^0 K_S^0\pi^+\pi^-$ six-track candidates. A clear signal of the 
correlated production of the pair of charged $K^*(892)$'s is seen (crosses), 
in agreement with the expected distribution for the simulated 
$e^+e^-\to K^*(892)^+K^*(892)^-$ reaction (dots). 
Figure~\ref{masses1}(b) shows the projection plot of (a) (four entries per 
event) which we fit with the sum of a double-Gaussian distribution for the 
$K^*(892)$ signal and a polynomial function for the background. 
The background includes also wrongly assigned $K_S^0\pi$ combinations. 
Gaussian parameters are taken from the simulated histogram shown in 
Fig.~\ref{masses1}(b). The fit yields 788$\pm$73$\pm$95 events: a second 
uncertainty is from a variation of the fit functions. Each event from the 
$K^*(892)^+K^*(892)^-$ reaction contributes twice, so a half of this value 
should be compared to $596\pm27$, the total number of the six-track signal 
events.  The obtained number indicates that the contribution of the 
$K^*(892)^+K^*(892)^-$ intermediate state does not exceed $66\pm11$\%.

With our data we cannot quantitatively extract a contribution from
production of a single $K^*(892)$ or from events without 
$K^*(892)$'s.

We calculate the background-subtracted invariant mass for
the two pions in the six-track sample, not originating from $K_S^0$, 
shown in Fig.~\ref{masses2}(a), 
and for the $K_S^0\pi^+\pi^-$ invariant mass (two entries per event), shown 
in Fig.~\ref{masses2}(b). 
These distributions indicate that  the $\rho(770)$ 
resonance in the $\pi^+\pi^-$ invariant mass, and   the $K_1(1270)$
resonance in the  $K_S^0\pi^+\pi^-$ invariant mass distribution cannot 
be excluded.
The solid histograms show the simulated distributions of the 
$K^*(892)^+K^*(892)^-$ intermediate state summed with the 30\% 
contribution from the $K_1(1270) K_S^0\to K_S^0K_S^0\rho(770) $ intermediate 
state. The latter contribution is shown by the dotted histogram.
 With the current data sample we cannot quantitatively extract this 
contribution. 

\section{Detection efficiency}
\label{sec:efficiency}
\hspace*{\parindent}

In our experiment, the acceptance of the DC for charged tracks is not 
100\%, and the detection efficiency depends on the production dynamics of 
the reaction as well as on the track reconstruction efficiency in the DC.

\begin{center}
\begin{figure}[tbh]
\vspace{-0.2cm}
\includegraphics[width=1.0\textwidth]{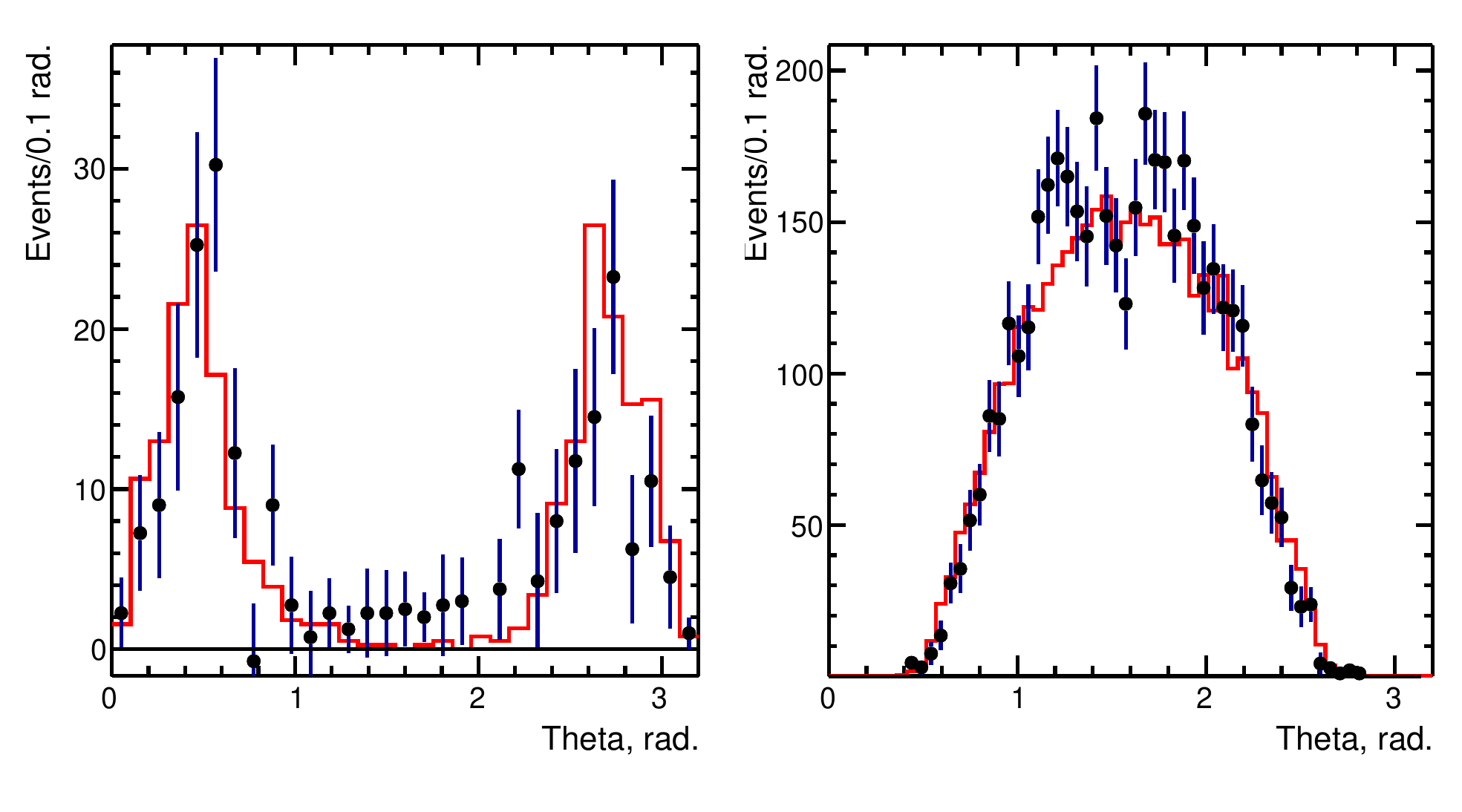}
\put(-245,173){\makebox(0,0)[lb]{\bf(a)}}
\put(-45,173){\makebox(0,0)[lb]{\bf(b)}}
\vspace{-0.8cm}
\caption
{(a) The background-subtracted experimental (dots) polar angle distribution  
in comparison with the simulated distribution (histograms) for the missing 
pion (a) and all detected pions (b). 
}
\label{5/6tracks}
\end{figure}
\end{center}

\begin{center}
\begin{figure}[tbh]
\vspace{-0.cm}
\includegraphics[width=1.\textwidth]{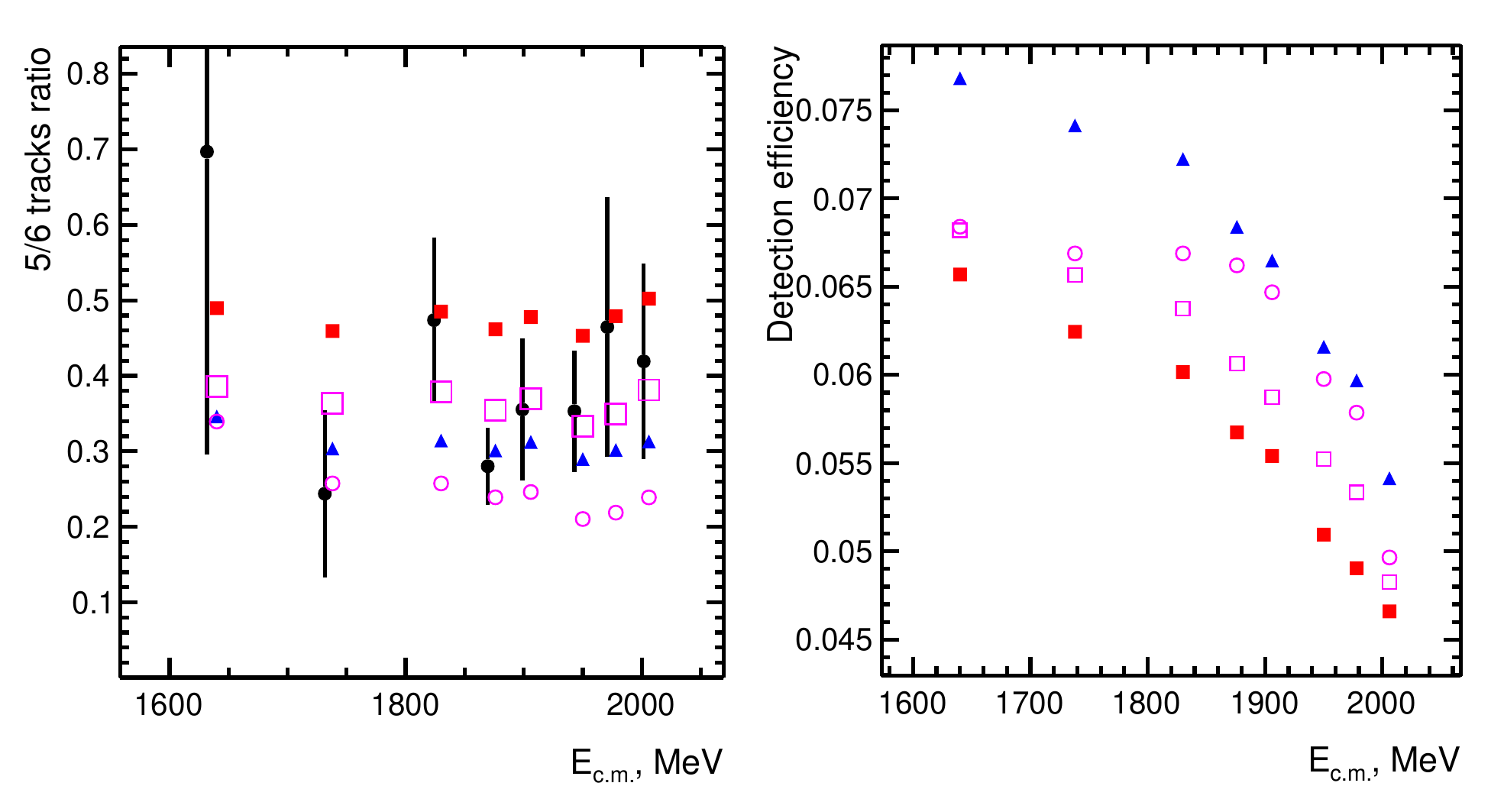}
\put(-245,163){\makebox(0,0)[lb]{\bf(a)}}
\put(-45,163){\makebox(0,0)[lb]{\bf(b)}}
\vspace{-0.5cm}
\caption
{ (a) The ratio of the number of five- to six-track events for data (dots) 
and simulation for the different intermediate states: phase space
model (squares), 
$K^*(892)^+ K^*(892)^-$ intermediate state 
(open squares), $K_1(1400)K_S^0$ (triangles), and $K_1(1270)K_S^0$ 
intermediate state (open circles).
(b) Detection efficiency obtained from the MC simulation for the 
$e^+ e^-\to K_S^0 K_S^0\pi^+\pi^-$ reaction for the different
intermediate states (symbols legend is the same as for (a)).
}
\label{efficiency}
\end{figure}
\end{center}
To obtain the detection efficiency, we simulate $K_S^0 K_S^0\pi^+\pi^-$ 
production in the primary generators, 50000 events for each c.m. energy 
interval for each model, 
pass simulated events through the 
CMD-3 detector using the GEANT4~\cite{geant4} package, and reconstruct them
with the same software as experimental data. 
We calculate the detection efficiency from the MC-simulated events
as a ratio of events after the selections described in 
Secs.~\ref{select}, \ref{dynamics} to the total number of generated events. 

Our selection of six- and five-track signal events allows us to estimate a 
difference in the tracking efficiency in data and simulation.
Figure~\ref{5/6tracks} shows by dots the background-subtracted polar angle 
for a missing pion (a) and for all detected pions (b). The histogram represents
the simulated distribution for the $K^*(892)^+ K^*(892)^-$ intermediate state. 
We observe reasonable agreement for data and simulation in these distributions
as well as in the calculated ratio of the number of five- to six-track events 
at each c.m. energy interval, shown in Fig.~\ref{efficiency} 
(a) by open squares.
The values of the ratio for the phase-space model (squares), 
$K_1(1400)K_S^0$ (triangles), and $K_1(1270)K_S^0$ intermediate state 
(open circles) are also shown in Fig.~\ref{efficiency}(a) and they are 
less compatible with data. 

We calculate the detection efficiency for the sum of 
events with six and five detected tracks.
Figure~\ref{efficiency}(b) shows the detection efficiencies obtained for 
the $e^+ e^-\to K_S^0 K_S^0\pi^+\pi^-$ reaction for different intermediate 
states: markers are the same as for Fig.~\ref{efficiency} (a). The detection 
efficiencies for different modes  are relatively close, and the 
efficiencies calculated 
for the $K^*(892)^+ K^*(892)^-$ intermediate state only and calculated with 
a 30\% contribution from the $K_1(1270)K_S^0$ reaction differ by less
or about 5\%.     
\begin{center}
\begin{figure}[tbh]
\vspace{-0.2cm}
\includegraphics[width=1.0\textwidth]{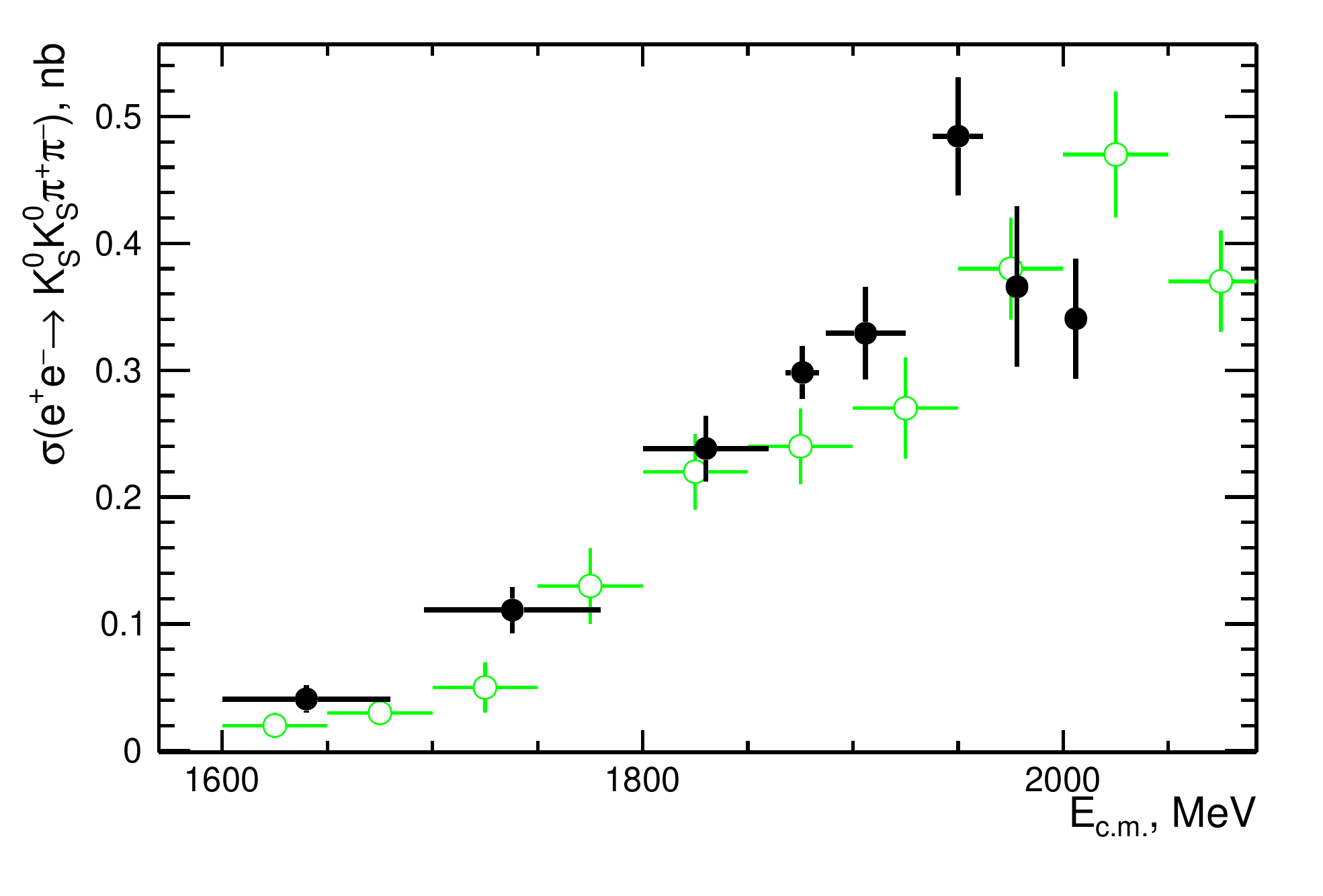}
\vspace{-1.cm}
\caption{
The $e^+e^-\to K_S^0 K_S^0\pi^+\pi^-$ cross section measured with the CMD-3
detector at VEPP-2000 (circles). 
The results of the BaBar 
measurement~\cite{isrksks2pi} are shown by open circles. 
}
\label{cross}
\end{figure}
\end{center}
\section{Cross section calculation}    
\hspace*{\parindent}
In each energy interval the cross section is calculated as 
$$
\sigma = \frac{N_{6\pi}+N_{5\pi}}{L\cdot\epsilon\cdot(1+\delta)},
$$ 
where $N_{6\pi}$, $N_{5\pi}$ are the background-subtracted numbers of 
signal events with six and five tracks, $L$ is the integrated
luminosity for this energy interval, 
$\epsilon$ is the detection efficiency, 
and $(1+\delta)$ is the radiative correction calculated 
according to Ref.~\cite{kur_fad,actis}.
To calculate the radiative correction, we use BaBar data for the 
$e^+e^-\to K_S^0 K_S^0\pi^+\pi^-$ reaction ~\cite{isrksks2pi} as a first 
approximation, and obtain $(1+\delta) = 0.92$ with very weak energy 
dependence.  

We calculate the cross sections for the $e^+e^-\to K_S^0 K_S^0\pi^+\pi^-$ 
reactions using the efficiency shown by open squares in 
Fig.~\ref{efficiency}(b)  for the\\  $K^*(892)^+ K^*(892)^-$ intermediate 
state. The cross section is shown in Fig.~\ref{cross}. We also calculate 
the cross section by using only events with six detected tracks: 
a less than 5\% difference is observed.

The energy interval, integrated luminosity, the number of six- and five-track 
events, efficiency, and the obtained cross section for each energy interval 
are listed in Table~\ref{xs_all}. 
\section{Systematic uncertainties}
\hspace*{\parindent}
The following sources of systematic uncertainties are considered.

\begin{itemize}

\item{The tracking efficiency was studied in detail in our previous 
papers~\cite{cmd4pi,cmd6pi}, and the correction for the track reconstruction 
efficiency compared to the MC simulation is about 1.0$\pm$1.0\% per track.
Since we add events with one missing track (from the two not from $K^0_S$),  
the MC-simulated detection efficiency is corrected by (--5$\pm$3)\%: 
the uncertainty is taken as the corresponding systematic uncertainty. 
}
\item {
The model dependence of the acceptance is determined by comparing 
efficiencies calculated for the different production 
dynamics. The maximum difference of the detection efficiencies of the 
dominant $K^*(892)^+ K^*(892)^-$ intermediate state and those for
other states is about 15\%. a possible admixture (of about 30\%) of
other states changes the efficiency by
about 5\%, what is taken as the systematic uncertainty estimate. 
}  
\item{
Since only one charged track is 
sufficient for a trigger (98--99\% single track efficiency), and using
a cross check with the independent neutral trigger, we conclude 
that for the multitrack events the 
trigger inefficiency gives a negligible contribution to the systematic 
uncertainty. 
}
\item{
The systematic uncertainty due to the selection criteria is studied by 
varying the requirements described above and doesn't exceed 5\%. 
}
\item{
The uncertainty on the determination of the integrated luminosity 
comes from the selection criteria of Bhabha events, radiative
corrections and calibrations of DC and CsI and does not exceed 
1\%~\cite{lum}.
}
\item{
The uncertainty in the background subtraction is studied
by the variation of the tile dimensions (20$\times$20, 25$\times$25, 
and 30$\times$30 MeV/c$^2$ 
dimensions tested), and by the comparison of the cross section 
calculated using only six-track events. A less than 5\% difference is observed.
}
\item{
The radiative correction uncertainty is estimated as about 
2\%, mainly due to the uncertainty on the maximum allowed energy of the 
emitted photon, as well as from the uncertainty on the cross section.
}
\end{itemize}

The above systematic uncertainties summed in quadrature give an overall
systematic error of about 10\%.

\begin{table}[tbh]
\caption{Energy interval, integrated luminosity, number of signal 6-track events,  number of signal 5-track events, 
detection efficiency, and the obtained cross section for the $e^+e^-\to K_S^0 K_S^0\pi^+\pi^-$ reaction.
Only statistical uncertainties are shown. 
}
\label{xs_all}
\vspace{-0.cm}
\begin{center}
\renewcommand{\arraystretch}{0.85}
\begin{tabular}{cccccccc}
\hline
{E$_{\rm c.m.}$, MeV} & {L, nb$^{-1}$} 
&{$N_{6\pi}$}
&{$N_{5\pi}$}
&{$\epsilon$}
&{$\sigma_{K_S^0 K_S^0\pi^+\pi^-}$, nb}\\

\hline
2007.0$\pm$0.5 & 4259  & 45  $\pm$ 7  & 19.0 $\pm$ 5.0  & 0.048 & 0.341 $\pm$ 0.047\\ 
1980$\pm$1     & 2368  & 29  $\pm$ 6  & 13.5 $\pm$ 4.1  & 0.053 & 0.366 $\pm$ 0.063\\ 
1940--1962     & 5230  & 95  $\pm$ 10 & 33.8 $\pm$ 6.7  & 0.055 & 0.484 $\pm$ 0.047\\ 
1890--1925     & 5497  & 72  $\pm$ 9  & 25.8 $\pm$ 5.9  & 0.059 & 0.329 $\pm$ 0.037\\ 
1870--1884     & 16803 & 218 $\pm$ 17 & 61.5 $\pm$ 10.1 & 0.061 & 0.298 $\pm$ 0.021\\ 
1800--1860     & 8287  & 79  $\pm$ 11 & 37.2 $\pm$ 7.0  & 0.064 & 0.238 $\pm$ 0.026\\ 
1700--1780     & 8728  & 47  $\pm$ 8  & 11.5 $\pm$ 4.8  & 0.066 & 0.111 $\pm$ 0.018\\ 
1600--1680     & 7299  & 11  $\pm$ 4  & 7.8  $\pm$ 3.7  & 0.068 & 0.041 $\pm$ 0.011\\
\hline
\end{tabular}
\end{center}
\end{table}

\section*{ \boldmath Conclusion}
\hspace*{\parindent}
The total cross section of the process $e^+e^-\to K_S^0 K_S^0\pi^+\pi^-$ 
has been measured using 56.7 pb$^{-1}$ of integrated 
luminosity collected by the CMD-3 detector at the VEPP-2000 $e^+e^-$ collider
in the 1.6--2.0 GeV c.m. energy range. The systematic uncertainty is 
about 10\%. 
From our study  we can conclude that the observed cross section can be 
described by the 
$e^+e^-\to K^*(892)^+ K^*(892)^-$ reaction, but an about 30--35\% contribution 
of the $K_1(1270)K_S^0$ intermediate state is not excluded.
The measured cross section for the $e^+e^-\to K_S^0 K_S^0\pi^+\pi^-$ reaction  
agrees with the only available measurement by BaBar~\cite{isrksks2pi}.

\subsection*{Acknowledgments}
\hspace*{\parindent}
The authors are grateful to A.I.~Milstein
for his help with theoretical interpretation and development of
the models. 
We thank the VEPP-2000 team for excellent machine operation. 
The work is partially supported by the Russian 
Foundation for Basic Research grant 17-02-00897.

\end{document}